\documentclass[breaklinks=true,colorlinks=true,citecolor=blue]{aa}  

\usepackage{graphicx}
\usepackage{orcidlink}
\usepackage{txfonts}
\usepackage{linenoaa}
\linenumbers
\renewcommand{\apjl}{ApJL}

\begin{document} 

   \title{An automated, self-calibration based pipeline for high-fidelity solar imaging with LOFAR: \textsc{SIMPL}}

\author{
  S.~Dey\inst{1}\thanks{Corresponding author: sohamd943@gmail.com}\orcidlink{0009-0006-3517-2031}
  \and
  D.~Oberoi\inst{1}\orcidlink{0000-0002-4768-9058}
  \and
  P.~Zucca\inst{2}\orcidlink{0000-0002-6760-797X}
  \and
  M.~Mancini\inst{2}\orcidlink{0000-0002-3861-9234}
  \and
  D.~Patra\inst{1}\orcidlink{0009-0003-8133-6621}
  \and
  D.~Kansabanik\inst{3,4}\orcidlink{0000-0001-8801-9635}
}

   \institute{National Centre for Radio Astrophysics, Tata Institute of Fundamental Research,
              S. P. Pune University Campus, Pune 411007, India
         \and
             ASTRON – The Netherlands Institute for Radio Astronomy, Oude Hoogeveensedijk 4, 7991 PD Dwingeloo, The Netherlands
        \and
            NASA Jack Eddy Fellow, University Corporation for Atmospheric Research, 3090 Center Green Dr, Boulder, CO, USA 80301
        \and
            The Johns Hopkins University Applied Physics Laboratory, 11001 Johns Hopkins Rd, Laurel, USA 20723
             }

  \abstract
   {The LOw Frequency ARray (LOFAR) is capable of imaging spectroscopy of the Sun in the 10–240 MHz frequency range, with high spectral, temporal, and spatial resolution. However, the complex and rapidly varying nature of solar radio emission -- spanning several orders of magnitude in brightness further exacerbated by the strong ionospheric phase distortions during daytime observations, poses major challenges for calibration, imaging, and automation.}
   {We aim to develop a fully automated, high-fidelity imaging pipeline optimized for LOFAR solar observations, capable of handling the intrinsic variability of solar emission and producing science-ready images with minimal human intervention.}
   {We have built the Solar Imaging Pipeline for LOFAR (SIMPL), which integrates excision of radio-frequency interference (RFI) for the solar-specific scenarios, calibration strategies, and self-calibration. The pipeline is designed to enable scalable and uniform processing of large archival datasets.}
   {SIMPL achieves more than an order-of-magnitude improvement in imaging dynamic range compared to previous efforts and reliably produces high-quality spectroscopic snapshot images. It has been tested across a wide range of solar conditions. It is currently being employed to process a decade of LOFAR solar observations, providing science-ready FITS images for the community and enabling both comprehensive and novel studies of solar radio phenomena -- ranging from quiet Sun emission and faint non-thermal features to active regions and their associated dynamic events, such as transient bursts.} 
   {}

    \keywords{Sun: corona --
             Radio interferometry
    }
   \maketitle
%

\section{Introduction}\label{sec:intro}
Imaging the radio Sun with high fidelity is an intrinsically challenging endeavour. 
The solar corona is a complex, dynamic and magnetized plasma environment that produces radio emissions across a broad range of angular scales and emission mechanisms. 
These emission mechanisms give rise to brightness temperature ($T_B$) that can vary many orders of magnitude over small time and frequency intervals \citep[e.g.,][]{kansabanik2022workingprinciple} that span a vast range of angular scales. 
At meter wavelengths, for instance, thermal bremsstrahlung from the quiescent Sun typically shows $T_B \sim 10^6$ K \citep[e.g.,][]{mercier2015,oberoi2017,vocks2018}. 
In stark contrast, coherent emission processes, often triggered by flare-accelerated electron beams, can generate extreme type III bursts with $T_B$ soaring beyond $10^{11}$ K \citep[e.g.,][]{melrose1989, saint-hilaire2013}. 
Capturing these emissions covering an immense dynamic range is further complicated by their rapid spectro-temporal evolution on sub-second timescales with fractional bandwidths $\delta\nu/\nu_0 \ll 1$ \citep[e.g.,][]{ratcliffe2014, clarkson2023}. 
Consequently, achieving a comprehensive understanding of coronal dynamics necessitates high-fidelity spectroscopic snapshot imaging capable of ``freezing" these rapid changes in both time and frequency.

The challenge is exacerbated by the frequent superposition of diverse morphological structures, ranging from intensely bright, compact bursts to faint, diffuse background emissions \citep[e.g.,][]{kansabanik2022workingprinciple, dey2025}.
Furthermore, the Sun's extreme flux density compared to typical calibrator sources introduces significant complexities in achieving robust calibration \citep[e.g.,][]{Mondal2019aircars,kansabanik2022workingprinciple,kansabanik2022paircarsalgorithm}.
Historically, these limitations, compounded by the capabilities of earlier instruments, meant that radio imaging studies of the dynamic solar emission were often restricted to the brightest emission sources at a few discrete frequencies \citep[e.g. see numerous examples in][]{mcleanlabrum1985,Pick2008, nindos2020}. 
Investigations of the much fainter quiet Sun, conversely, usually necessitated integration times spanning many hours \citep[e.g.,][]{mercier2009,zhang2022}. 
Even with an advanced instrument like the LOw Frequency ARray (LOFAR; \citet{vanhaarlem2013}), imaging studies often focused on the brightest events or relied on long integrations.

The advent of new-generation metric radio interferometers, including LOFAR (operating in the 10–90 MHz and 110–240 MHz bands), the Long Wavelength Array at the Owens Valley Radio Observatory (OVRO-LWA; \citet{anderson2018ApJ, hallinan2023}; 12-85 MHz), and the Murchison Widefield Array (MWA; \citet{tingay2013, wayth2018}; 80–300 MHz), driven by enormous advances in digital signal processing and computational power, represents a significant leap in instrumental capabilities. 
These instruments are exceptionally well-suited for solar imaging with high temporal and spectral resolution at low radio frequencies and are already yielding a wealth of interesting scientific results \citep[e.g.,][]{morosan2015,morosan2019,suresh2017,zucca2018,zucca2025,mccauley2019,mohan2019,mondal2020,mondal2020cme,mondal2023, chhabra2021,sharma2022,bhunia2023, oberoi2023, kansabanik2024cme,dey2025}. 
Dynamic imaging spectroscopy with LOFAR, for example, has begun to unveil previously unseen burst morphologies, such as spatially resolved radio signatures of electron beams in coronal shocks \citep{zhang2024} and images of intricate fine structures seen in dynamic spectra associated with type III bursts \citep{dabrowski2025}. 
With more than a decade of such LOFAR observations —- more than 2200 hours spanning Solar Cycles 24–25 now archived, the sheer volume of this information-rich data makes manual analysis impractical. 
A fully automated and robust pipeline is therefore essential for the uniform processing of this extensive dataset and for enabling crucial long-term statistical studies.

In this work, we present {\it Solar IMaging Pipeline for LOFAR} (SIMPL) -- a fully automated and robust calibration and imaging pipeline specifically designed for solar radio observations with LOFAR. 
SIMPL integrates solar-specific Radio Frequency Interference (RFI) mitigation techniques, tailored calibration strategies, and a self-calibration framework optimized for the Sun's complex and dynamic emission. 
Together, these enable the generation of high-fidelity spectroscopic snapshot images, significantly improving the dynamic range and fidelity of LOFAR solar radio images. 
The pipeline is currently being employed to systematically process over a decade of solar interferometric observations, to provide science-ready FITS files that can be accessed on demand by the broader solar physics community.  

The paper is organized as follows. Section \ref{sec:lofar} provides a brief overview of the LOFAR instrument and its solar observing modes. 
Section \ref{sec:challenges} outlines the challenges in calibration and imaging of solar observations at low radio frequencies. 
In Section \ref{sec:prevpipelines}, we review existing solar imaging pipelines for LOFAR and their limitations. 
Section \ref{sec:algdsc} presents the algorithmic design of SIMPL. 
Results demonstrating the pipeline's performance are shown in Section \ref{sec:results}. 
Section \ref{sec:hpc} describes the integration of the pipeline to the existing LOFAR processing framework, and finally, Section \ref{sec:futurework} outlines directions for future development, followed by conclusions in Section \ref{sec:conclusion}.

\section{LOFAR}\label{sec:lofar}
LOFAR is a state-of-the-art interferometer operating across the 10-240 MHz frequency range. It comprises 38 stations in the Netherlands and 14 international stations across Europe. Out of the 38 Dutch stations, 24 stations are located within $\sim$2 km at Exloo. This cluster of stations is referred to as the ``core''. Each station hosts two co-located antenna fields -- Low-Band Antennas (LBA; 10--90 MHz) and High-Band Antennas (HBA; 110--240 MHz) - each providing data in both orthogonal linear polarizations, with signals digitized and beamformed locally before being streamed to the central correlator in Groningen.  In the interferometric mode \footnote{\href{https://science.astron.nl/telescopes/lofar/lofar-system-overview/technical-specification/frequency-subband-selection-and-rfi}{LOFAR system overview}}, LOFAR provides the following specifications: 
\begin{itemize}
    \item The instantaneous bandwidth is split into 512 subbands via a polyphase filter.
    \item Up to 244 subbands ($\approx 48$ MHz total bandwidth) can be selected and sent to  the correlator in 16-bit mode (or 488 subbands in 8-bit mode).
    \item Each subband spans 195.3 kHz or 156.2 KHz (for the 200 MHz or 160 MHz clock respectively), which can be further channelized (e.g., into 64 channels yielding $\sim 3$ kHz resolution). 
    \item Minimum integration time is of $\sim 0.16$ seconds.
\end{itemize} 

\begin{figure*}
    \centering
    \includegraphics[width=\linewidth]{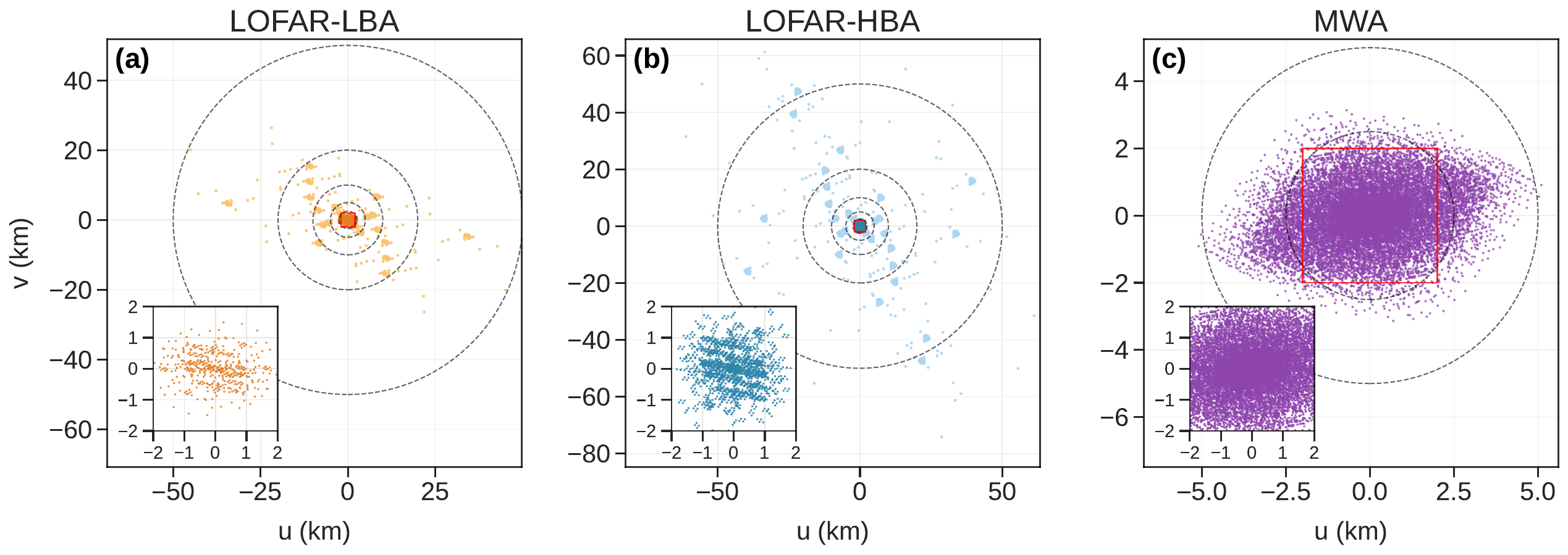}
    \caption{\textbf{Instantaneous \textit{uv-}coverage comparison between LOFAR and MWA}. 
    Projected baseline distributions for: (a) LOFAR-LBA, (b) LOFAR-HBA, and (c) MWA Phase-II configuration. 
    Both LOFAR-LBA and HBA exhibit densely clustered cores, offering high sensitivity to large-scale solar structures, complemented by longer remote baselines that enable higher spatial resolution. 
    In contrast, the MWA provides a significantly denser and more uniformly distributed \textit{uv-}coverage within a compact footprint, facilitating high-fidelity snapshot imaging at moderate resolution (a few arcminutes). 
    Insets in each panel zoom into the central $\pm 2$ km region (marked by red squares) to highlight the baseline distributions within the core. 
    The dashed black circles are drawn with radii of 2.5, 5, 10, 20 and 50 km. 
    The number of independent baselines from core antennas is: 276 for LOFAR-LBA and 1128 for LOFAR-HBA. 
    For panel (c), representing the MWA phase-II configuration, there are 8128 instantaneous baselines within 5 km.}  
    
    \label{fig:uv-coverage_comparison}
\end{figure*}

LOFAR allows concurrent observations of multiple sky regions \citep{broekema2018} using simultaneous station beams. 
For solar observations, a strong calibrator source (e.g., Cassiopeia A or Taurus A) is usually observed simultaneously using an independent LOFAR station beam. 
This approach avoids interrupting the solar observation for separate calibrator scans, as is done conventionally, and instead derives complex antenna gains from the simultaneously observed calibrator, which are then applied to the solar beam.
Considering only the core stations, 276 baselines are available instantaneously for the LBA and 1128 for the HBA. 
Instantaneous uv-coverage for a single spectral slice for LOFAR is shown in Figure \ref{fig:uv-coverage_comparison}: LBA (a) and HBA (b). 
Thus, LOFAR provides high time and spectral resolution along with reasonable instantaneous uv-coverage, giving it spectroscopic snapshot imaging capability necessary for studying narrowband and transient solar emissions.

\section{Calibration challenges}\label{sec:challenges}

\begin{figure}
    \centering
    \includegraphics[width=\linewidth]{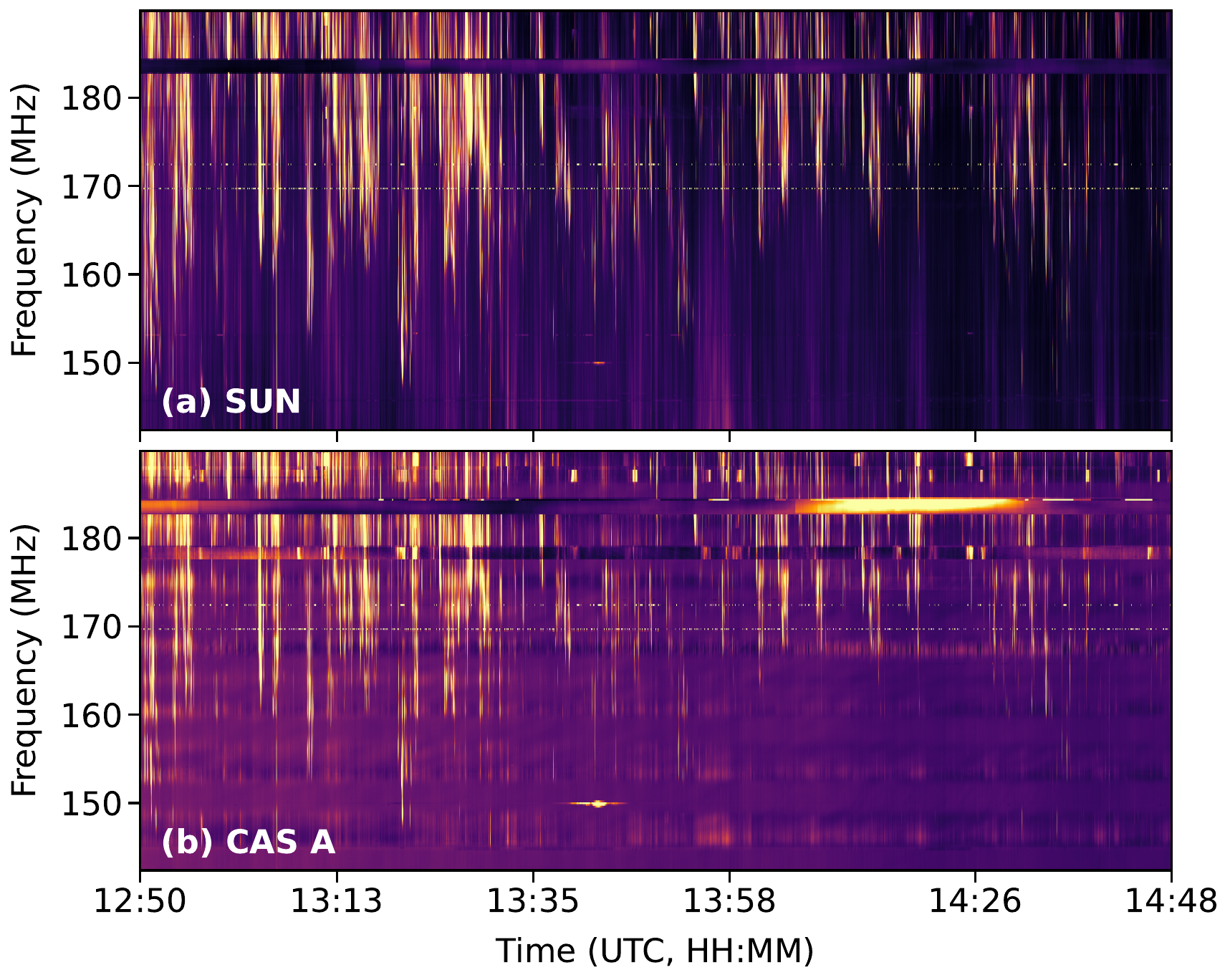}
    \caption{\textbf{Comparison of dynamic spectra (DS) from the Sun (a) and the calibrator Cassiopeia A (b), observed simultaneously with two separate LOFAR station beams.} 
    These DS were generated by averaging visibilities in the $150$-$500~ \lambda$ range in the \textit{uv}-plane to preferentially capture transient, compact structures while suppressing contributions from large-scale quiet Sun emission.
    Although Cassiopeia A was located $11.4^\circ$ from the Sun during the observation, its DS shows features that strongly resemble those seen in the solar spectrum, indicating substantial solar contamination in the calibrator field. 
    This highlights the difficulty of obtaining clean calibrator data during solar observing sessions, particularly with wide FoV instruments like LOFAR.}
    \label{fig:comparison_solar_cal_DS}
\end{figure}

As mentioned in Section \ref{sec:intro}, the dynamic nature and extremely high flux densities of solar radio emission present substantial challenges for its calibration. 
The strong solar signals, including during quiet times, can contaminate calibrator observations, even when the calibrator is located several degrees away from the Sun. 
This issue is especially pronounced for wide Field of View (FoV) aperture array instruments like LOFAR. 
Figure \ref{fig:comparison_solar_cal_DS} illustrates this problem: the upper panel shows dynamic spectra of the Sun, while the lower panel shows that for the simultaneously observed calibrator source using a separate station beam, Cassiopeia A, located $11.4$ degrees away. 
Signatures of essentially all of the narrow-band short-lived solar emissions are evident in the dynamic spectra of the calibrator beam, demonstrating significant contamination of the solar signal into the calibrator beam. 
While the contaminations of the time/frequency variable component of the solar emission are easily seen, those of the quiet Sun are not visually evident from the dynamic spectrum.

Minimizing this contamination by observing a calibrator much farther away introduces a different major calibration difficulty at low radio frequencies, arising from ionospheric effects. 
The ionosphere introduces spatially and temporally varying phases across the sky, resulting in direction dependent differential phase errors across the array. 
For narrow FoV interferometers, this is typically addressed by interleaving observations of the target source with frequent scans of a nearby phase calibrator.
However, for solar observations, and with wide FoV instruments, - where the Sun overwhelmingly dominates the sky emission - this approach becomes impractical.
With LOFAR, an ``A-team source" is observed as a calibrator. 
These are the extremely bright radio sources in the northern sky - Cassiopeia A, Cygnus A, Taurus A, and Virgo A.
These observations capture the sum of direction-independent instrumental phases and direction-dependent ionospheric phases toward the calibrator. 
However, the derived phase solutions cannot correct for the (often large) differential phase toward the direction of the Sun, and this uncorrected phase difference can be large enough to significantly degrade image quality.

\begin{figure*}
    \centering
    \includegraphics[width=\linewidth]{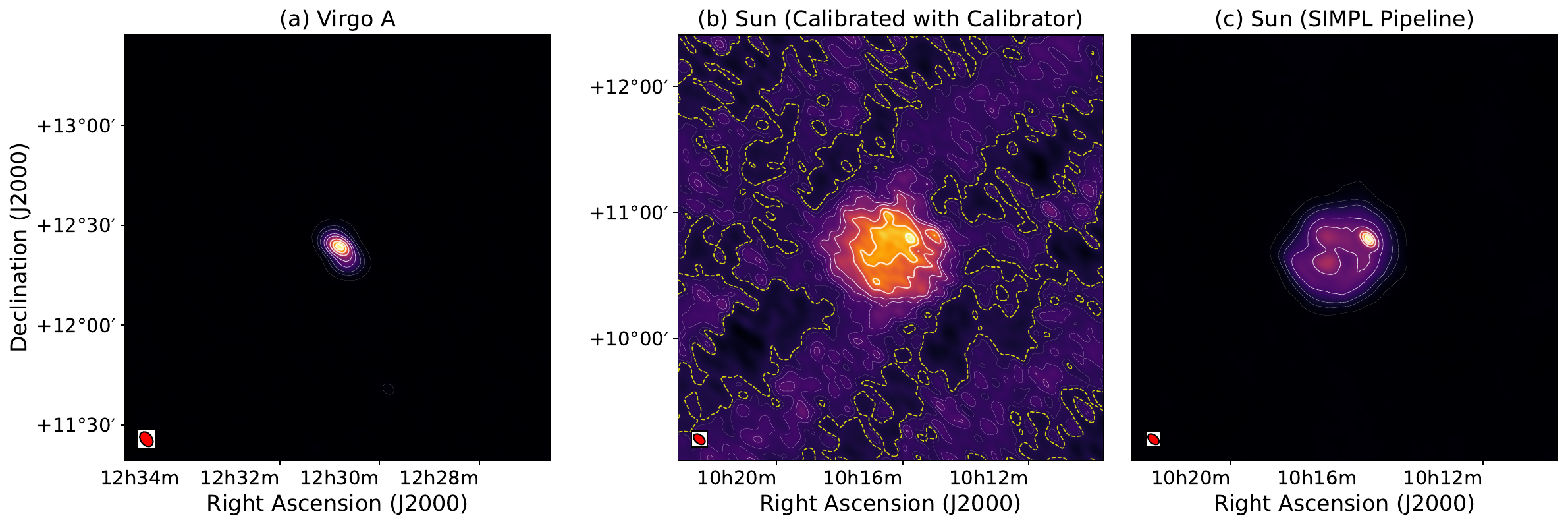}
    \caption{\textbf{Demonstration of the limitations of applying calibrator-based phase solutions across widely separated sky regions.} Left panel: Image of Virgo A using the calibration solutions obtained from the calibrator beam, demonstrating excellent fidelity and dynamic range exceeding 1250 (rms: 0.08 Jy/beam).
    Middle panel: Solar image obtained by applying calibration solutions derived from the calibrator source Virgo A. The image suffers from significant phase errors, resulting in distorted morphology and a limited dynamic range of $\sim28$ (rms: 4.0 Jy/beam). 
    Right panel: Solar image produced by the \textsc{SIMPL} pipeline using a self-calibration-based approach, yielding a markedly improved dynamic range of $\sim550$ (rms: 0.5 Jy/beam). 
    Contours at each panel are drawn at $-1\%$, $1\%$, $5\%$, $10\%$, $20\%$, $30\%$, $50\%$, $70\%$ and $90\%$ of their respective peaks, with the negative contour shown as yellow dashed lines.
    The red ellipse in the bottom-left corner of each panel indicates the restoring beam.
    The comparison underscores the necessity of direction-dependent calibration for accurate high-fidelity solar imaging.}
    \label{fig:calib_with_calibrator}
\end{figure*}

Figure \ref{fig:calib_with_calibrator}  demonstrates this issue. The left panel shows an image of the calibrator (Virgo A) at a centre frequency of 113 MHz, made using a time and frequency integration of 53 s and 2 MHz, respectively. 
The image has a dynamic range exceeding 1250 and an rms of 0.08 Jy/beam. 
The flux density of Virgo A at 150 MHz is 1209 Jy \citep{degasperin2020}, which is comparable to the quiet Sun flux density at this frequency.
The middle panel shows the solar image obtained after applying calibration solutions derived from Virgo A at the same central frequency and shorter spectral and temporal averaging spans of 1 MHz and 2 s, respectively.
Accounting for the differences in spectral and temporal averaging, the expected rms noise level in the solar image -- estimated based on the measured rms from the Virgo A -- is 0.58 Jy/beam. 
The achieved map rms is 4.0 Jy/beam, resulting in an imaging dynamic range of $\sim28$. 
Note that beam gain corrections have not been applied to these images.
The right panel shows the image for the same data and time frequency integration made by SIMPL, which achieves a map rms of 0.5 Jy/beam, leading to an imaging dynamic range exceeding 550. 
The imaging strategy employed by SIMPL is detailed in Section \ref{sec:algdsc}.

\section{Limitations of the existing LOFAR solar imaging pipelines}\label{sec:prevpipelines}

The calibration challenges described in Sections \ref{sec:intro} and \ref{sec:challenges} have long posed significant hurdles for solar imaging, including some specific to LOFAR. 
Two notable efforts are commonly used for solar imaging studies with LOFAR - (i) the LOFAR Solar Imaging Pipeline developed by the Leibniz-Institut für Astrophysik Potsdam (AIP) group \citep{breitling2015} and (ii) LOFAR-Sun\footnote{\url{https://lofar-sun-tools.readthedocs.io/en/latest/}} \citep{zhang2022}, a more recent and modular Python-based pipeline. 
Both of them build on standard tools from the LOFAR imaging ecosystem and rely on the calibration solutions from a calibrator source, usually observed simultaneously with the Sun. 
No solar-specific adaptations are made in their calibration strategies. 
Their processing workflow largely follows the procedures used in LOFAR's standard imaging pipeline\footnote{\href{https://support.astron.nl/confluence/spaces/public/pages/144349257/LOFAR+Data+Reduction+reference+manual}{LOFAR data reduction reference manual}} \citep{vanhaarlem2013}, with minor modifications such as skipping steps like automated flagging and source finding. 
In some cases, calibrator solutions have been manually derived during quiet solar periods when no active emission was evident.

As discussed in Section \ref{sec:challenges}, these exclusively calibrator-based approaches can often compromise the solar imaging quality, particularly under conditions of elevated solar activity or significant ionospheric variability.
Although \citet{breitling2015} mentions optional phase only self-calibration in their workflow, to the best of our knowledge, it has not been used in subsequent studies employing their pipeline \citep[e.g.,][]{dabrowski2025, brose2025A&A}.
This underscores the need for dedicated solar calibration strategies tailored to the unique challenges of low-frequency solar radio interferometry. 

\section{Algorithm description}\label{sec:algdsc}
SIMPL incorporates the learnings and the experience gained from the development of dedicated solar imaging pipelines for the MWA \citep{Mondal2019aircars, kansabanik2022paircarsalgorithm,2023ApJSpipelineimplementation}, which introduced self-calibration strategies well-suited for compact, centrally condensed arrays. 
The solar flux density usually drops substantially at longer baselines. 
Consequently, SIMPL is currently tailored to process solar interferometric data using only the LOFAR core stations and ignores the sparsely sampled remote baselines. 

Unlike the MWA, which typically operated with 128 tiles in its phase I \citep{tingay2013} and II \citep{wayth2018} (and now up to 256 tiles in its phase III), producing 8128 simultaneous independent baselines within 5 km, LOFAR's compact core provides significantly fewer baselines -- 276 for the LBA and 1128 for the HBA within 2 km (Figure \ref{fig:uv-coverage_comparison}).
The sparser \textit{uv}-coverage provided by LOFAR, in comparison with the MWA, inherently limits the achievable spectroscopic, snapshot imaging dynamic range with LOFAR.
Additionally, LOFAR typically observes calibrator sources simultaneously with the Sun (usually one of the A-team sources), while MWA schedules separate nighttime calibrator observations. 

\begin{figure*}
    \centering
    \includegraphics[width=\linewidth, trim={0.9cm 0.7cm 2.5cm 0.7cm},clip]{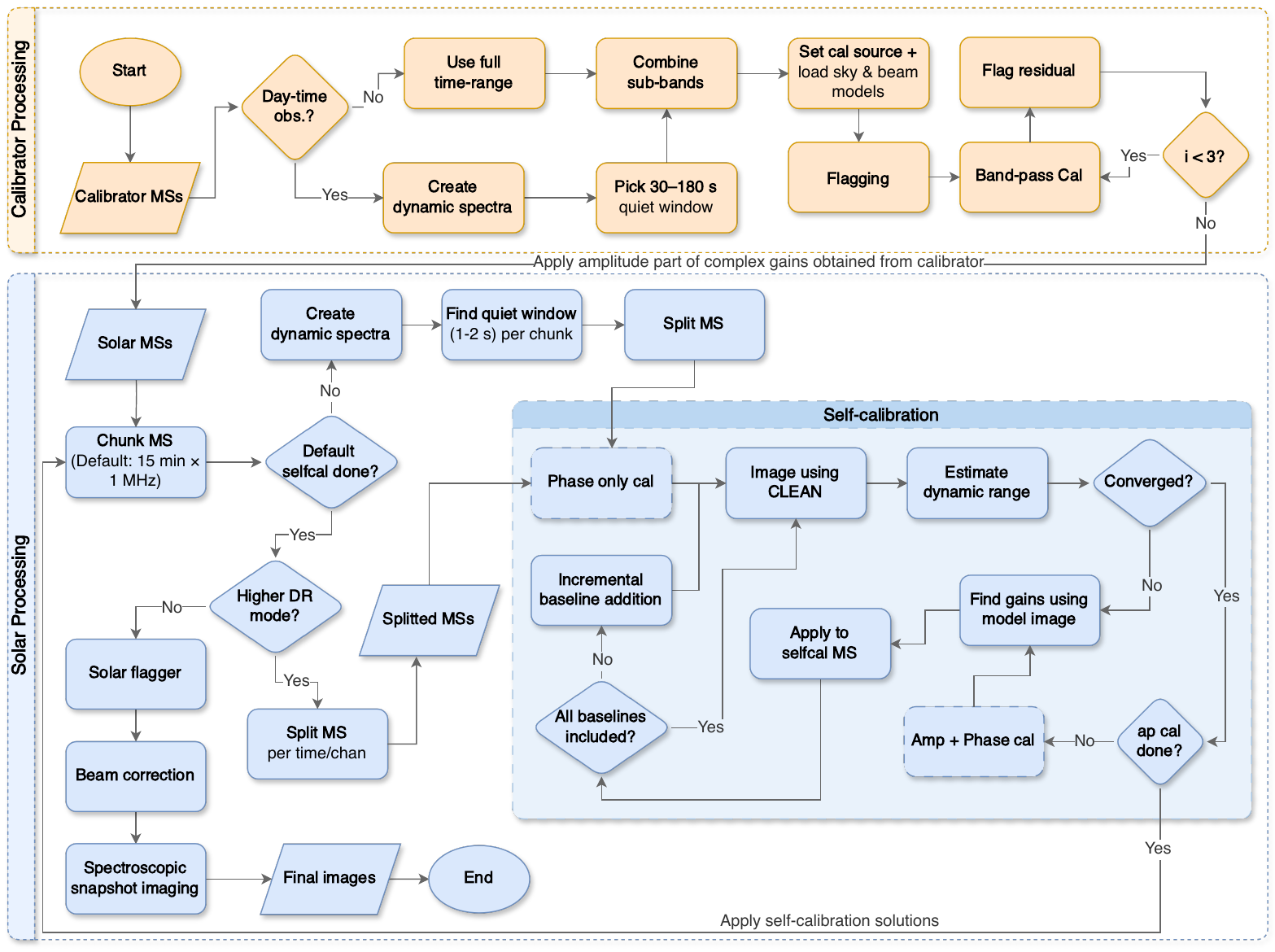}
    \caption{\textbf{Schematic overview of \textsc{SIMPL}.} The flow is divided into three main sections: calibrator processing (orange), solar processing (light blue), and the self-calibration sub-module (darker blue). 
    Calibrator processing identifies an optimal quiet time range for daytime observations, derives antenna gain solutions, and applies only the amplitude part of the gains to solar data. 
    Solar Processing operates on time–frequency chunks of the solar measurement set, invoking the self-calibration block to iteratively refine phase (and later amplitude) solutions.
    An optional higher dynamic range mode estimates differential self-calibration solutions per time and frequency slice.
    Final steps include primary beam correction and generation of spectroscopic snapshot images. 
    Rounded rectangles denote processing steps, diamonds represent decision points, and dashed rounded rectangles indicate iterative self-calibration loops.}
    \label{fig:pipeline_flowchart}
\end{figure*}

To provide an overview, in its current implementation, SIMPL first identifies the optimal time range within the calibrator observation to extract reliable calibration solutions. 
It then applies only the amplitude part of the antenna gain solutions to the solar data. 
The phase part of these solutions is entirely derived via self-calibration, using an iterative strategy with progressive baseline inclusion. 
The solar dataset is segmented into time-frequency chunks, and self-calibration is independently carried out and applied to each segment to account for the time and frequency variations of complex gains, dominated primarily by variations in ionospheric phases. 
Finally, primary beam correction for the direction of the Sun is applied, followed by the generation of spectroscopic snapshot images at user-defined time and frequency resolution and intervals. 
A schematic overview of the processing pipeline is illustrated in Figure \ref{fig:pipeline_flowchart}, and details are provided in the following sub-sections.

\subsection{Identifying optimal time for calibration from calibrator observation} \label{sec:optimal_time_calibration}

As demonstrated by the presence of solar emission features in the calibrator beam in Figure \ref{fig:comparison_solar_cal_DS}, simultaneous daytime calibrator observations can be severely contaminated by solar emission.
The dynamic spectra was generated by averaging visibilities in the $150$-$500~ \lambda$ range in the \textit{uv}-plane to preferentially capture transient, compact structures while suppressing contributions from large-scale quiet Sun emission.

To extract reliable calibration solutions from the daytime calibrator scan, it is essential to identify a time interval that is minimally affected by solar contamination. 
In SIMPL, we implement a composite-score-based algorithm to automatically detect the quietest time window within the calibrator dynamic spectrum. 

The algorithm is designed to identify a contiguous time segment with minimal statistical variation and minimal transient solar activity.
A user-configurable sliding window (SW; default: 30 s) is moved across the time axis using the available bandwidth, and the following statistical metrics are computed:
\begin{itemize}
    \item Standard deviation ($\sigma$) and median absolute deviation (MAD)
    \item Skewness and kurtosis: indicators of non-Gaussianity, with elevated values typically marking impulsive radio bursts
    \item  Shannon entropy: a measure of information content or randomness within the spectral profile, normalized to account for relative signal scaling.
\end{itemize}
Each metric is normalized independently using min-max scaling, where for a given metric $M$, the normalized value is computed as:
\begin{equation}
    M_{norm} = \frac{M-min(M)}{max(M) - min(M)}
\end{equation}
The composite score, $z$, is then calculated as:
\begin{equation}
    z=w_1*\sigma+w_2*\mathrm{MAD}+w_3*\mathrm{Skew}+w_4*\mathrm{Kurt}+w_5*(1-\mathrm{Entropy)}
\end{equation}
The weights $w_1$ to $w_5$ are empirically determined from extensive testing across a large number of datasets and are set to $0.4$, $0.4$, $0.1$, $0.1$, and $0.1$, respectively.
Higher $z$ values indicate stronger contamination.
The resulting composite score time series is smoothed using a moving median filter with a window length of $\textrm{SW}/3 + 1$.

\begin{figure}
    \centering
    \includegraphics[width=\linewidth, trim=0 0 0 0,clip]{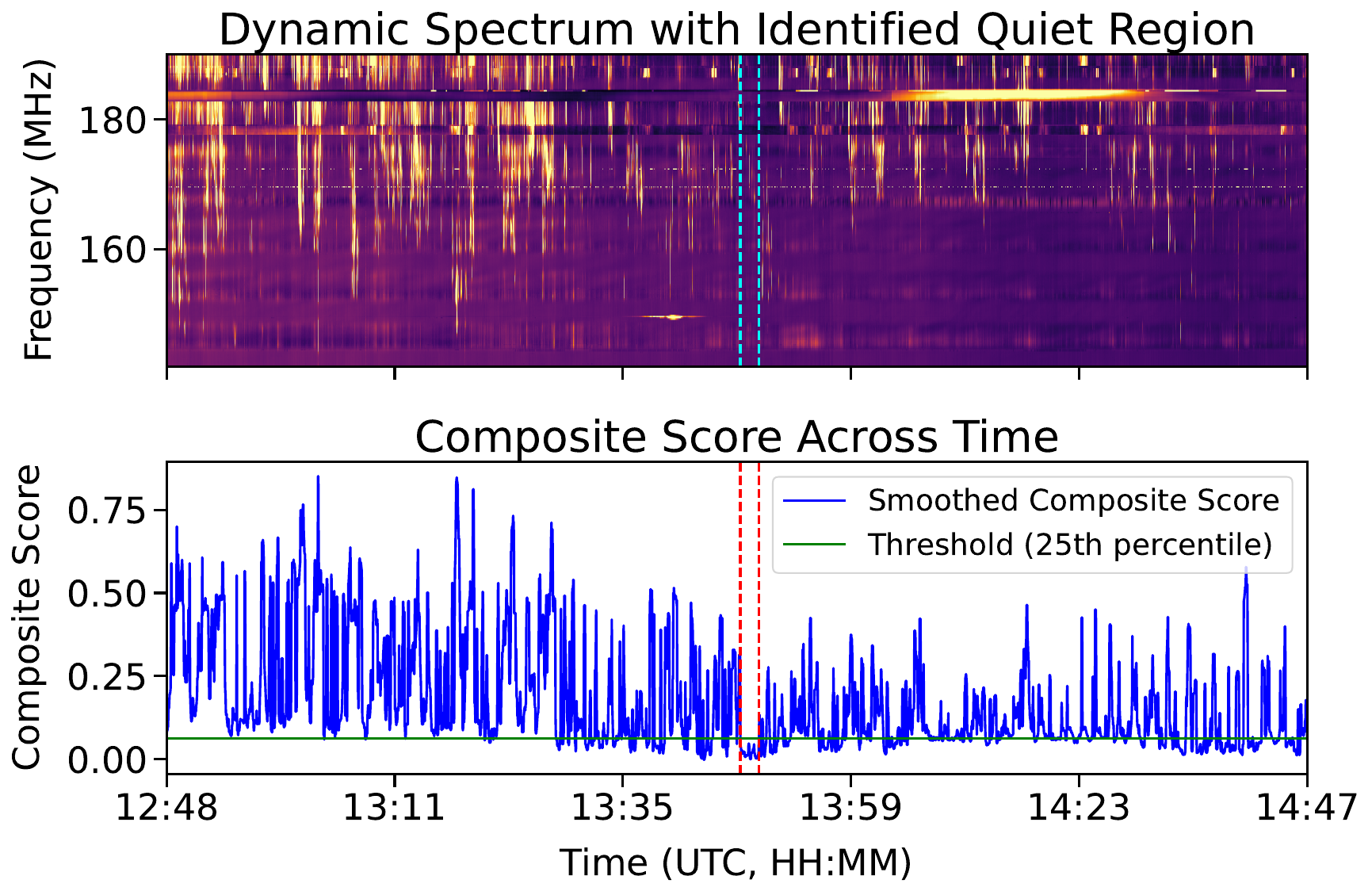}
    \caption{\textbf{Automated identification of quiet calibration interval within contaminated calibrator data.} The top panel shows the dynamic spectra of a calibrator (Cassiopeia A) observation contaminated by solar emission, with vertical dashed lines in cyan marking the start and end of the automatically identified quiet interval. The bottom panel displays the smoothed composite score time series, derived from statistical measures of variability, non-Gaussianity, and entropy across sliding time windows. The horizontal green line marks the 25th percentile threshold used to define "quiet" intervals. The selected segment (between red dashed lines, from 13:48:20 to 13:50:18 UTC) represents the longest continuous window below this threshold and is used for reliable calibration.}

    \label{fig:Calibrator_quiet_time}
\end{figure}

A quiet-time threshold is then defined as the 25th percentile of the smoothed composite score distribution.
All windows with $z$ below this threshold are considered as ``quiet", and the longest contiguous quiet segment is selected for calibration.
Figure~\ref{fig:Calibrator_quiet_time} illustrates this procedure and highlights an example of the identified quiet interval with the calibrator dynamic spectrum which was severely contaminated by transient solar flux.

\subsection{Calibration using simultaneous calibrator} \label{sec:simulcalib}

To mitigate solar contamination from calibrator observation, SIMPL first identifies a time interval during which the Sun is quiescent (see Section \ref{sec:optimal_time_calibration}). 
This ensures that the calibration solutions are estimated from a time range of calibrator scan, which is devoid of transient solar contamination. 
Additionally, to avoid the contamination due to the quiet Sun emission from the extended solar disc,  visibilities are restricted to an appropriate \textit{uv}-range (typically greater than $100 ~\lambda$).

The calibrator source is first identified and the corresponding model is imported using Default Preprocessing Pipeline (DP3) \citep{vandiepen2018ascl.soft04003} after multiplying with the LOFAR station beam. 
The raw calibrator dataset is initially flagged for RFI using the \textsf{tfcrop} algorithm in the \textsf{flagdata} task in CASA \citep{bean2022}. 
Calibration is then performed in three iterations using the CASA task \textsf{bandpass}. 
After each iteration, automated flagging is performed using the \textsf{rflag} algorithm on the residual visibilities to remove low-level RFIs, followed by recalibration to obtain updated gain solutions.

Due to the direction-dependence of ionospheric phases, the complex gain solutions derived from the direction of the calibrator are not the optimal solutions for the solar field.
Figure \ref{fig:calib_with_calibrator} shows an example where employing the gain solutions determined from the calibrator Virgo A yields a high fidelity image when applied to it, with  a dynamic range exceeding 1250 (Figure~\ref{fig:calib_with_calibrator}, left panel).  
The solar image obtained after applying the same calibration solutions suffers from significant phase errors and achieves a dynamic range of only $\sim28$ (Figure~\ref{fig:calib_with_calibrator}, middle panel), illustrating that ionospheric conditions differ substantially across different parts of the sky.

To overcome this, SIMPL retains only the amplitude component of the calibrator-derived gain solutions for absolute flux scaling and disregards the phase solutions. 
The phases are instead derived via self-calibration directly on the solar data as detailed in Section~\ref{sec:selfcal}. 
This approach significantly improves image quality: for the same example, a dynamic range of $\sim550$ is achieved (Figure~\ref{fig:calib_with_calibrator}, right panel), along with better morphological fidelity.

\subsection{Self-calibration} \label{sec:selfcal}

In standard radio interferometric calibration, where calibrator sources are used, the sky model is well known. The complex direction–independent
antenna gains, $g_{p}$, are determined by minimizing the following objective function:
\begin{equation}
  \Phi \;=\;
  \sum_{p,q}\left|\,V_{pq}-g_{p}g_{q}^{\ast}V^{\mathrm{M}}_{pq}\right|^{2},
  \label{eq:phi_known_model}
\end{equation}
where $V_{pq}$ is observed visibilities and $V^{\mathrm{M}}_{pq}$ is the model visibility, and the summation is over all baselines formed by antennas $p$ and $q$. 
Here $\ast$ denotes the complex conjugate, so $g_{q}^{\ast}$ is the complex conjugate of the gain $g_q$.
Equation~(\ref{eq:phi_known_model}) presumes that the sky brightness
distribution is accurately known \textit{a priori}.  
When this assumption fails, one may treat both the gains and the sky model
as free parameters and Equation
(\ref{eq:phi_known_model}) generalizes to
\begin{equation}
  \Phi \;=\;
  \sum_{p,q}\left|\,V_{pq}-g_{p}g_{q}^{\ast}\,
      \mathcal{F}\!\left[I^{\mathrm{M}}\right]_{pq}\right|^{2},
  \label{eq:phi_unknown_model}
\end{equation}
where $\mathcal{F}$ denotes the Fourier transform operator and $I^{\mathrm{M}}$ is the \textit{a priori} unknown sky brightness distribution.
The system can be solved as long as the number of free parameters describing the sky and the complex antenna gains remains smaller than the number of measured visibilities.
In practice, this approximation holds and the number of measured visibilities is generally much larger than the number of free parameters.
This permits the following iterative loop to be executed, minimizing $\Phi$, until convergence:
\begin{enumerate}
  \item[\textbf{1.}] \textbf{Imaging.}  Apply the best available gain solution to
        the visibilities and perform an inverse Fourier transform to obtain
        an image of the sky. 
  \item[\textbf{2.}] \textbf{Model update.}  Construct an updated model
        image $I^{\mathrm{M}}$, using a CLEAN based deconvolution algorithm (here using \texttt{WSClean} \citep{offringa2014}). 
  \item[\textbf{3.}] \textbf{Convergence test.}  Quantify the improvement
        in image (e.g.\ through the dynamic range).  If the change falls below a chosen threshold, exit the loop.
  \item[\textbf{4.}] \textbf{Gain solution.}  With the revised sky model
        fixed, minimize Equation~(\ref{eq:phi_unknown_model}) to derive a new set
        of complex gains $g_{p}$.
  \item[\textbf{5.}] \textbf{Repeat.}  Return to step~\textbf{1} with the
        updated gains and iterate until convergence.
\end{enumerate}

\begin{figure*}
    \centering
    \includegraphics[width=0.9\linewidth]{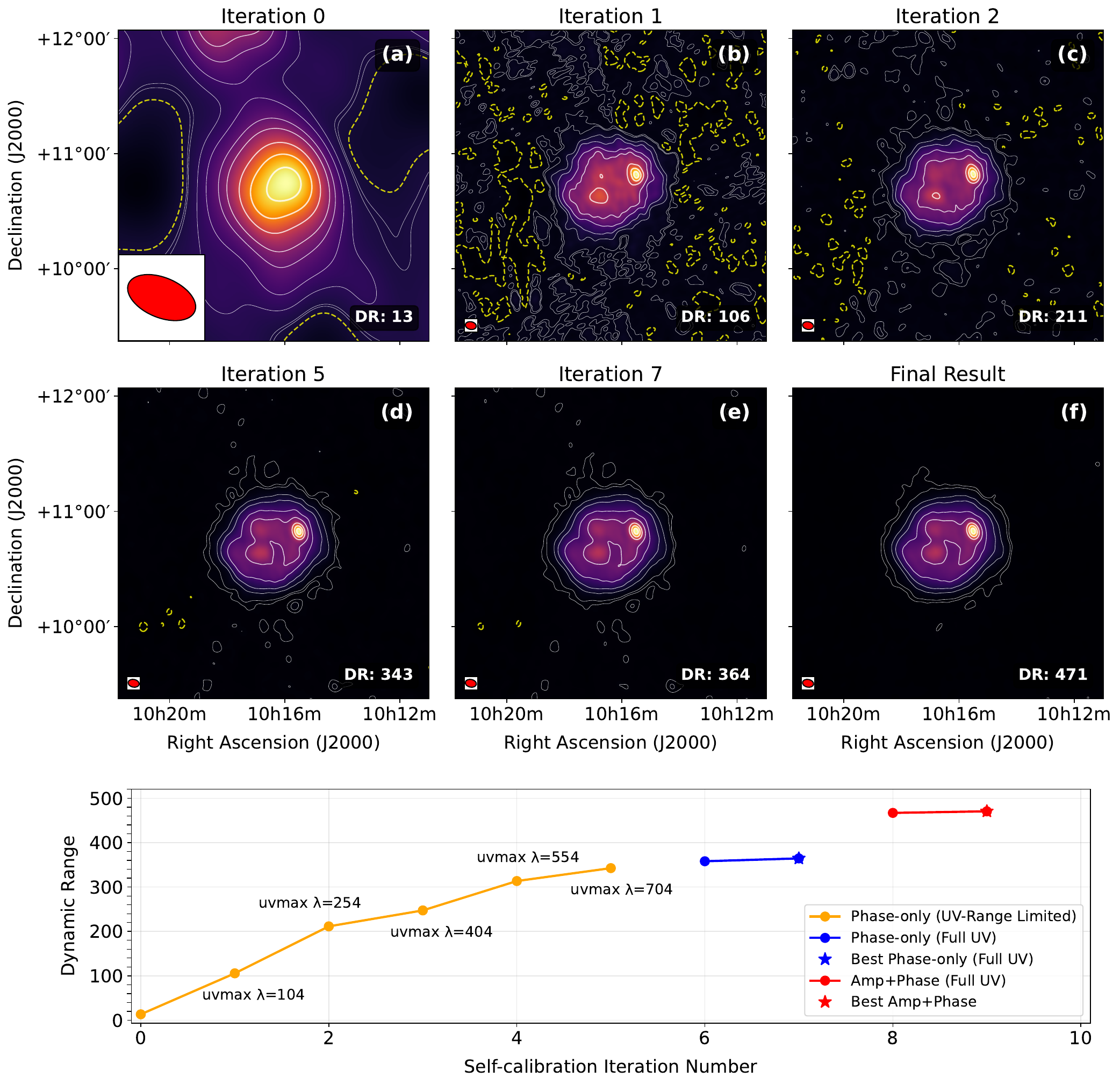}
    \caption{\textbf{Dynamic range and image quality improvement through self-calibration.} 
    Top two rows (a--f): Improvement of the imaging quality over multiple self-calibration iterations, demonstrated at 112.6 MHz with 1 s time integration and 1 MHz frequency averaging. 
    White contour are at 0.5\%, 1\%, 5\%, 10\%, 20\%, 30\%, 50\%, 70\% and 90\% of the peak intensity. 
    The yellow dashed contour represents $-1\%$ of the peak. 
    The red ellipse in the bottom-left corner of each panel indicates the restoring beam. 
    (a) Initial image (iteration 0), after applying only amplitude solutions from the calibrator and visibilities restricted to $uv < 104~\lambda$, allowing a simple Gaussian model to initiate phase-only self-calibration. 
    (b) Image after first phase-only self-calibration using the Gaussian model; dynamic range (DR) improves to 106. 
    (c) Phase-only self-calibration extended to $uv < 254~\lambda$ for the second iteration further improves DR to 211. 
    (d) After including baselines up to $uv < 704~\lambda$ at iteration 5, DR increases to 343. 
    (e) Final phase-only iteration with DR = 364. 
    (f) Final image after amplitude-phase self-calibration converges; DR reaches 471. 
    Bottom panel: Dynamic range variations across self-calibration cycles. 
    Each point represents an iteration, and the maximum baseline length used ($uv_\mathrm{max}$) is mentioned in the figure. 
    Phase-only calibration steps are shown in blue/yellow, amplitude-phase steps in red. 
    Star markers indicate the best dynamic range achieved in each stage. 
    The improvement in image fidelity through iterative self-calibration is evident.}

    \label{fig:dr_improv}
\end{figure*}

The foregoing strategy -- often referred to as self-calibration \citep{conrwell1981, pearson1984} -- is now standard in improving the dynamic range of radio interferometric images.  
Because the objective function is nonlinear and non-convex, the algorithm can stagnate in a local minimum if the initial sky model is poor and/or the data are noisy. 
Providing a realistic starting model is therefore essential to guide the optimization toward the physically correct solution. 
Since in our case we are not applying phases of the solutions obtained from calibrator observation, we use a simple Gaussian model at the phase centre as our initial model. 
To ensure that this model is consistent with observation, we do the first iteration using only the baselines for which the Sun remains unresolved. 
In subsequent iterations, we add baselines with larger \textit{uv-}distances in steps until the phase-only self-calibration converges, following a similar convergence strategy employed by \citet{kansabanik2022workingprinciple,kansabanik2022paircarsalgorithm}. 
Only when the phase solutions for all antennas are reasonably well constrained, amplitude-phase self-calibration is performed. 
Figure \ref{fig:dr_improv} shows an example of dynamic range improvement during self-calibration.

\subsubsection{Alignment of solar disk using quiet solar time}\label{sec:alignment}

At low radio frequencies, ionospheric electron density variations cause the apparent position of the Sun to drift over time and frequency. 
Self-calibration, while effective in correcting relative gain errors across the array, can not provide absolute phase information. 
This inherently loses absolute positional information on the sky. 
When self-calibration is applied under such conditions, the combined effect can result in significant misregistration of the Sun in the image plane.

To address this, self-calibration is performed on a short (1–2 s) duration of data during a quiet solar period, when the solar disk is clearly visible. 
A Gaussian model placed at the phase center is used in the initial iteration to ensure that the solar center aligns with the phase center. 
This step helps preserve the correct solar position, allowing the resulting images to be accurately transformed to the helioprojective coordinate system, essential for comparing solar radio images to those obtained at other wavebands.
The ability to image the solar disk, and hence the identification of a quiet interval, is critical to correct registration of the radio image: 
if only a compact feature located away from the disk center is visible, this method will incorrectly align that feature to the phase center.
The quiet-time selection follows the procedure in Section~\ref{sec:optimal_time_calibration}, but applied to a maximum time span of 2 s.

\subsection{Flagging solar observations}
\label{subsec:flagging}

\begin{figure*}[h!]
    \centering
    \includegraphics[width=\linewidth]{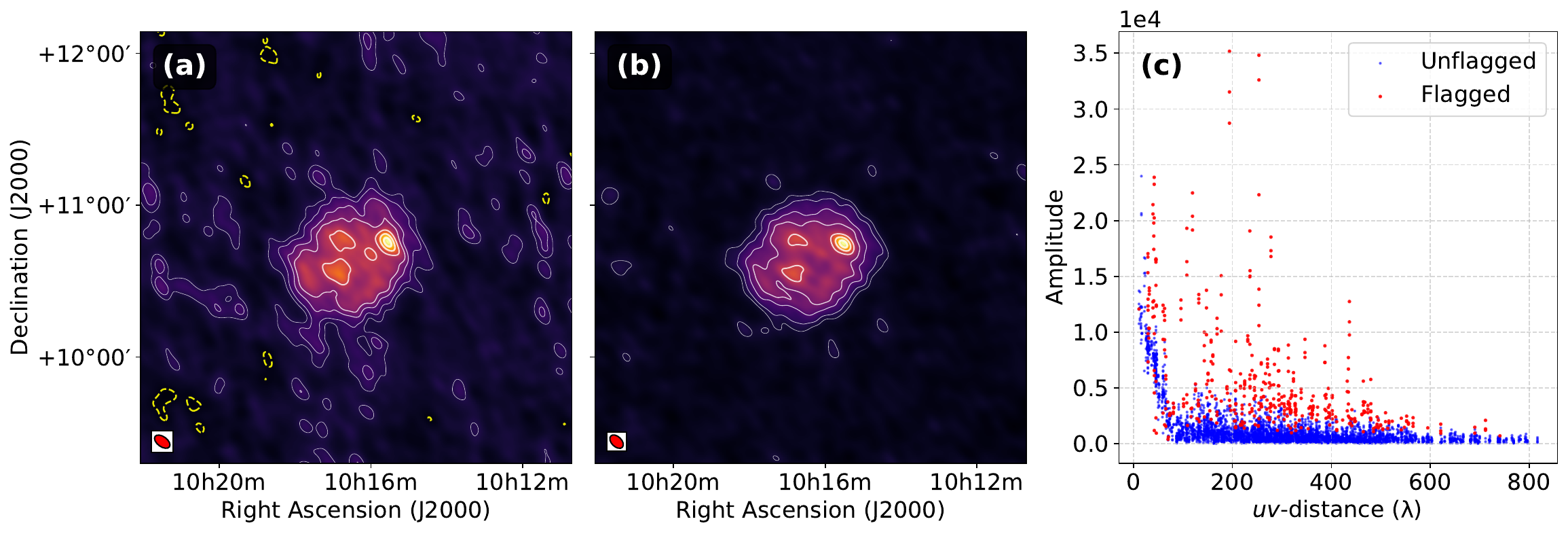}
    \caption{\textbf{\textit{uv-}domain based flagging on solar visibilites.}
    Panels (a) and (b) show solar images at the same frequency and time, before and after SIMPL’s uv-based flagging, respectively. 
    These are at 111.13 MHz with 1 s time integration and 195.3 kHz averaged in frequency and are severely affected due to RFI. 
    For panels (a) and (b) white contours are drawn at 5\%, 10\%, 20\%, 30\%, 50\%, 70\% and 90\% of the peak intensity and yellow dashed contour represents the negative 5\% of the peak intensity.
    The restoring beam is show in red in bottom left corners.
    The image (a) exhibits an rms of 13.7 Jy/beam and a dynamic range of 37, while the image (b), obtained after outlier flagging followed by another round of self-calibration, shows an improved rms of 7.5 Jy/beam and a dynamic range of 74.
    Panel (c) illustrates the corresponding visibility amplitudes as a function of \textit{uv-}distance. 
    Flagged visibilities (red) were identified as outliers as detailed in Section \ref{subsec:flagging}. 
    This approach enables effective suppression of corrupted visibilities while preserving genuine solar structure across spatial scales.}
    \label{fig:sflag}
\end{figure*}

Flagging solar radio datasets poses a unique challenge: automated flagging algorithms that rely on detecting statistical outliers in time and frequency often misidentify genuine solar emissions — such as solar radio bursts — as RFI. 
This is because solar radio bursts are statistically indistinguishable from RFI in time-frequency phase space. 
To address this issue, SIMPL implements a flagging strategy based on identifying outliers in the \textit{uv-}domain independently for each time and frequency slice. 
This approach is effective even for transient solar radio emissions and has been implemented in a flagger called {\it aNKflag} \citep{2023ApJSpipelineimplementation}.
For ease of integration, SIMPL uses a custom implementation of a flagger inspired by {\it aNKflag}. It is optimised for computational efficiency and minimising over-flagging, and more details are provided next.

The Sun is a spatially complex and time-variable source, which results in significant variations in visibility amplitudes across the \textit{uv-}plane. 
However, the dominant trend is that the amplitudes at shorter \textit{uv-}distances (corresponding to large angular scales) are significantly higher than those at longer \textit{uv-}distances.
To ensure meaningful statistical analysis, we perform logarithmic binning in \textit{uv-}distance. This binning scheme uses finer bins at shorter \textit{uv-}ranges — where visibility amplitudes are generally higher and vary rapidly — and coarser bins at longer \textit{uv-}ranges, aligning well with the natural distribution of data density in the \textit{uv-}plane. 
The flagging algorithm implemented in SIMPL treats the real and imaginary components of complex visibilities as independent statistical populations. 
They are analysed separately for each time-frequency slice using identical \textit{uv-}binning and statistical analysis. 
Within each \textit{uv} bin, robust statistical outlier detection is performed using Median Absolute Deviation (MAD) with a deliberately lenient adaptive threshold. 
This threshold is typically $3-5\sigma$, depending on the number of data points in the bin, and progressively increases for bins with fewer data points, e.g., $7.5\sigma$ for bins containing fewer than 20 data points. 
For a dataset $X = \{x_1, x_2,~...,~x_n\}$ in each \textit{uv-}bin, a deviation score $d$ is estimated as:
\begin{equation}
    d_i = \frac{|x_i - M|}{MAD} ,
\end{equation}
where M is the median($X$).
Points are flagged if $d$ for either the real or the imaginary component exceeds the threshold in its respective \textit{uv-}bin.
Additional safeguards to prevent over-flagging are implemented: if more than 40\% of points in any \textit{uv-}bin are identified to be flagged, the algorithm applies a more relaxed threshold ($2\times$ the original). 
If, despite this, the number of points to be flagged remains greater than 40\%, no data are flagged for that \textit{uv}-bin.
For bins with more than 30 data points, the median calculation uses a trimmed estimator (removing 5\% from each tail) to enhance robustness against contamination. 
Flagging is performed on the entire solar dataset, independently for each time, frequency and correlation (XX and YY), on the \texttt{CORRECTED\_DATA} column after self-calibration solutions have been applied, with parallel processing across multiple timestamps for faster processing.

This approach allows SIMPL to effectively flag outliers even in the presence of spectrally complex and time-variable active solar emissions often seen in the solar dynamic spectra. 
Figure \ref{fig:sflag} (c) shows an example of flagging based on MAD on visibilities, for a time and frequency slice severely affected by RFI. 
Panel (a) shows the image formed (with self-calibration done) before this flagging was performed. 
The image has an rms of 13.7 Jy/beam and a dynamic range of 37. 
Panel (b) shows the image obtained after flagging was done, followed by self-calibration again. 
The resulting image has an rms of 7.5 Jy/beam and a dynamic range of 74, which is a factor of $\sim2$ improvement.

\subsection{Spectroscopic snapshot imaging of solar emissions}
After all the above steps are completed, the visibilities are corrected for the beam gains in the direction of the Sun.
This is done using the \texttt{everybeam}\footnote{\url{https://everybeam.readthedocs.io/en/latest/}} module integrated within the DP3 framework.
Following beam correction, spectroscopic snapshot imaging is performed using \texttt{WSClean} \citep{offringa2014}. 
By default, the imaging is done using a Briggs weighting scheme with a robustness parameter of 0.5, though this can be adjusted by the user depending on the desired trade-off between sensitivity and resolution. 
Other \texttt{WSClean} parameters have been empirically optimized based on extensive testing across a variety of LOFAR solar datasets.
Imaging is carried out at user-defined time and frequency intervals and integrations, enabling the generation of high-fidelity spectroscopic snapshot images of dynamic solar emission.

\subsection{Optional higher dynamic range mode}
\label{subsec:hdr}

In its default configuration, SIMPL partitions the measurement set into user-defined time and frequency chunks. 
Self-calibration solutions are computed independently for each chunk (using 1-2 s from within that chunk) and applied uniformly within it.
As discussed in Section \ref{sec:alignment}, a quiet time is preferred for this as it allows us to correct for ionospheric refraction, and working with a longer time chunk increases the likelihood of including a suitably quiet interval.
This approach offers ease of implementation, is computationally efficient, and, under most circumstances, delivers an imaging quality superior to that provided by earlier attempts.
However, during periods of high ionospheric activity, the calibration solutions can evolve significantly within an individual time chunk, reducing the gains from this approach. 
This effect is naturally more pronounced for LBA than for HBA.
Under such circumstances, reducing the time chunk size can lead to improved calibration accuracy and, consequently, better imaging quality.

Ionospheric activity tends to be higher during periods of higher solar activity.
Periods of high solar activity require one to look across longer time chunks to find instances of acceptably low solar activity.
On the other hand, dealing with the faster ionospheric variations demands working with shorter time chunks.
SIMPL deals with this by first computing self-calibration solutions from the quietest time data available in a suitably large time chunk and applying them to the entire chunk to compute the refractive shifts (Section \ref{sec:selfcal}).
This is followed by estimating self-calibration solutions at each time and frequency slice, as an optional step.
As the initial self-calibration already provides a reliable starting model, this differential self-calibration stage bypasses Gaussian modelling and progressive baseline inclusion, and starts with phase-only self-calibration, followed by amplitude and phase calibration till convergence is achieved.
While attractive from an imaging quality perspective, this approach tends to be challenging from a computational load perspective.

\begin{figure*}
    \centering
    \includegraphics[width=0.9\linewidth]{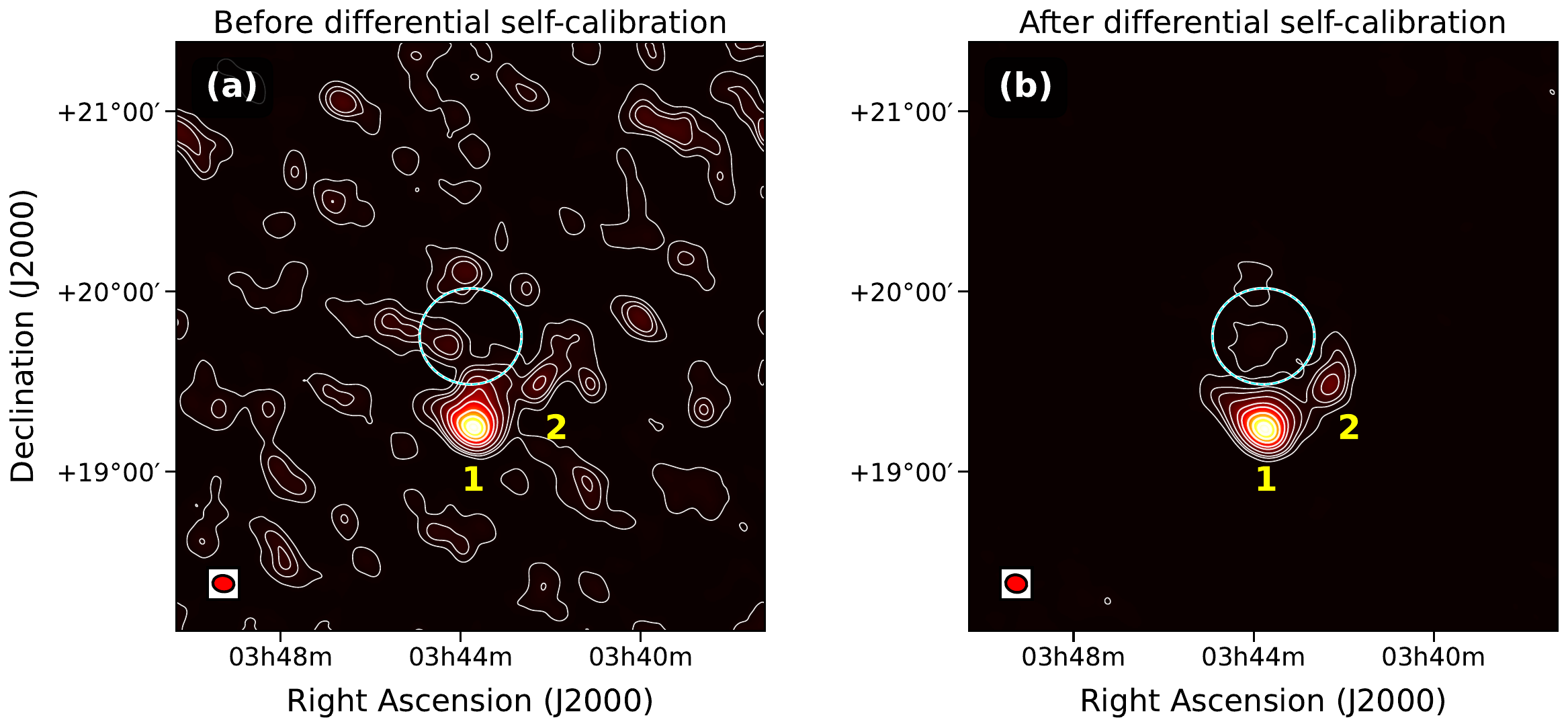}
    \caption{\textbf{Demonstration of SIMPL's optional higher dynamic range mode on a LBA dataset.} (a) Image at 77 MHz, obtained with 195.3 kHz frequency averaging and 1 s time integration, after initial self-calibration on longer time and frequency intervals. 
    This is before differential self-calibration and has a dynamic range (DR) of 86. (b) Image at the same time and frequency after subsequent differential self-calibration, showing significant improvement in imaging fidelity with DR 467. 
    The cyan circle denotes the optical solar disk. 
    Contours are drawn at 1\%, 3\%, 5\%, 10\%, 20\%, 30\%, 50\%, 70\% and 90\% of the peak. 
    The restoring beam is shown in red in bottom left corners. 
    The enhanced DR has enabled reliable detection of the faint source (marked 2), which is an order of magnitude fainter than source 1.}
    \label{fig:differential_selfcal}
\end{figure*}

Figure \ref{fig:differential_selfcal} illustrates the improvement in imaging fidelity achieved through this two-step calibration process. 
Panel (a) shows the image obtained after the initial self-calibration using longer time and wider frequency chunks. 
While the dominant solar emission is well detected, the residual phase errors result in elevated noise and degrade the dynamic range. 
Panel (b) shows the same time and frequency slice after performing differential self-calibration at full time and frequency resolution. 
The dynamic range improves by more than a factor of five, and some previously undetectable fainter sources -- such as source 2 -- are now clearly visible.  
This demonstrates the utility of differential self-calibration in capturing the complexity of solar emission morphology in greater detail, even under dynamic ionospheric and solar conditions, without compromising the positional alignment established during initial calibration.

\begin{figure*}
    \centering
    \includegraphics[width=\linewidth]{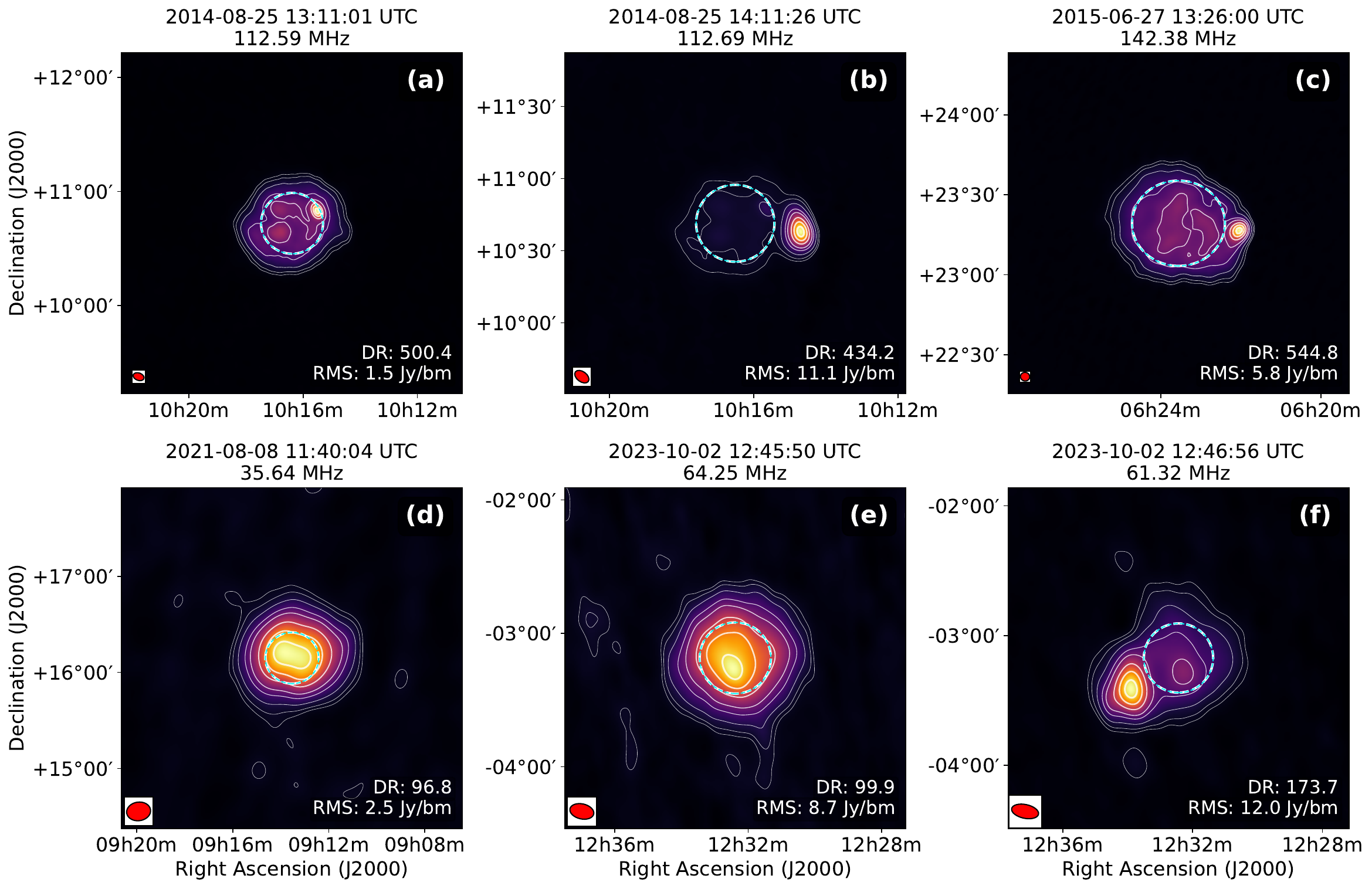}
    \caption{\textbf{Examples of spectroscopic snapshot images produced by SIMPL in its default mode.} 
    All images are generated using 1 s integration and 195.3 kHz frequency averaging, except for panel (d), which was integrated over 1 s and 1 MHz. 
    Contours are drawn at 3\%, 5\%, 10\%, 20\%, 30\%, 50\%, 70\% and 90\% of the maximum intensity in each image. 
    The red ellipse in the bottom-left corner of each panel indicates the restoring beam, and the cyan dashed circles represent the optical solar disk. 
    Top panels (a--c): HBA observations. (a) and (c) show examples of noise storm emission. 
    (b) displays a bright type III radio burst, more than an order of magnitude more intense than the noise storm in (a). 
    Low-level contours reveal the quiet Sun emission even in the presence of strong radio sources. 
    Bottom panels (d--f): LBA observations. 
    (d) shows a quiet Sun snapshot at 35.64 MHz with a dynamic range of 104. 
    (e) and (f) correspond to snapshots immediately before and during a type II burst. 
    The quiet Sun remains visible in the low-level contours, even in panel (f), despite the presence of an intense burst.}
    \label{fig:pipeline_output_demon}
\end{figure*}

\section{Results}\label{sec:results}

We have implemented the algorithm described in Section \ref{sec:algdsc}  in SIMPL and applied it to multiple LOFAR datasets spanning a range of solar conditions to evaluate its performance and robustness. 
SIMPL is currently configured to operate in two distinct modes: one optimized to maximize imaging quality without regard to computational efficiency, and the other that prioritizes faster runtimes by performing self-calibration on fewer time and frequency slices. 
In this section, we showcase results produced by the computationally efficient mode. 
Figure \ref{fig:pipeline_output_demon} presents a representative sample of the resulting spectroscopic snapshot images. 

Panels (a)--(c) show HBA observations, demonstrating SIMPL's ability to recover diverse solar emission features. In panels (a) and (c), compact sources associated with noise storm activity are visible, with surrounding contours revealing extended quiet Sun emission.
Panel (b) showcases a strong type III radio burst; here, the peak flux density exceeds the noise storm event shown in panel (a) by more than an order of magnitude. 
The pipeline recovers not only the intense type III burst, but also the noise storm region and the underlying quiet Sun structure. 
The dynamic range achieved for HBA datasets routinely exceeds 400 in quiescent conditions and can reach above 1000 during intense radio bursts. 
The resulting images show well-defined source morphology with minimal imaging artefacts, even in the presence of rapidly evolving emission. 

Panels (d)--(f) illustrate SIMPL's performance at lower frequencies using LBA data. 
Panel (d) shows a quiet Sun snapshot at 35.64 MHz (integrated over 1 s and 1 MHz), demonstrating a dynamic range of about 100, with the 3\% contour extending to $\sim 2.5~R_{\odot}$.
Snapshot imaging of extended emissions is significantly more challenging than compact sources, particularly with LOFAR LBA, which provides only 276 independent baselines considering only the core stations. 
Previous quiet Sun studies at these frequencies required 2--5 hours of integration \citep[e.g.,][]{zhang2022}. 
Panels (e) and (f) are 1 s, 195.3 kHz snapshot images capturing the onset of a type II radio burst, with panel (e) showing the quiet Sun emission before the event. 
Despite the bright and structured emission in panel (f), the quiet Sun remains visible in the lower contour levels. 
These examples highlight SIMPL's ability to observe low surface brightness features with high fidelity even in the presence of much brighter transient events.

Across both HBA and LBA observations, SIMPL consistently delivers high-quality spectroscopic snapshot images, achieving more than an order of magnitude improvement in dynamic range over what is possible when relying only on using calibration obtained towards the direction of the calibrator source -- the approach being used by earlier LOFAR imaging piplines (Figure \ref{fig:calib_with_calibrator}).
The imaging quality is maintained over a wide frequency range (35--190 MHz), varying solar activity levels, and across different observing epochs and solar conditions, underscoring the generality and robustness of the approach.


\section{Integration in the existing processing framework}\label{sec:hpc}
The LOFAR Solar and Space Weather key science project has recently undertaken the massive exercise to reprocess the archival solar observations from LOFAR.
Especially with the Incremental Development of LOFAR Space-weather \citep[IDOLS;][]{zhang2022AGU} observation campaign, which routinely gathers continuous solar and ionospheric observations, the growing number of observations necessarily require an automatic scheduling of data reduction pipeline to readily provide higher-level data products, without requiring any manual intervention. 
The framework used is called ATDB-LDV \citep{vermaas2019poster, bozkurt2024spie}, and it can orchestrate:  the data retrieval from the archive; the execution of the specified pipeline; and the final archival of processed data products. 
ATDB-LDV supports the execution of pipelines described in the Common Workflow Language (CWL\footnote{\url{https://www.commonwl.org/}}). 
This allows for flexible specification of pipeline parameters and execution strategies, as well as portability. 
Given the capability of SIMPL to calibrate both HBA and LBA interferometric datasets, we have implemented a CWL specification for the pipeline\footnote{\url{https://git.astron.nl/ssw-ksp/solar_imaging}} so that it can be automatically executed on the LOFAR Solar archive, and the final data products of images and video previews can be generated.

\section{Future development plans}\label{sec:futurework}

Although the current implementation of SIMPL is optimized for processing large datasets, several improvements are planned to enhance its performance and functionality. 
As described in Sections \ref{sec:selfcal} and \ref{subsec:hdr}, SIMPL currently offers two self-calibration modes, the default efficient mode, where a self-calibration solution is computed only for a single carefully chosen time-frequency patch per chunk and used for the whole of it, and an optional mode where an additional self-calibration solution is computed for every individual time and frequency slice being imaged.
The former is computationally lean and delivers good imaging quality except during periods of enhanced ionospheric activity, while the latter delivers improved imaging quality, especially under challenging ionospheric conditions, though at a much higher computational cost.
An optimal approach to provide enhanced imaging quality without incurring a very large computational cost is to compute self-calibration solutions over a user defined time-frequency grid fine enough to capture the ionospheric variations in antenna gains, but much coarser than the time-frequency integrations used for imaging. 
Suitably interpolated calibration solutions can then be applied to the entire dataset.
We plan to implement such a mode as the default in SIMPL.
This will allow us to reap much of the benefits from the per-slice calibration, but at a fraction of the associated computational cost.

We also plan to introduce two major new capabilities. While SIMPL represents a significant improvement over previous approaches, its current implementation does not fully exploit LOFAR's capability in two key respects.
First, though LOFAR can provide full polarimetric data, SIMPL is designed to produce only Stokes I (total intensity) images; and second, SIMPL uses only the LOFAR core stations, which lie within a 2 km region, and ignores the remote stations, which can provide baselines up to $\sim$100 km.
To extract the most information from the data, we intend to address both these limitations in future developments, as described in the following subsections. 

\subsection{Polarization calibration}
Full-Stokes calibration at low radio frequencies is challenging for several reasons. Firstly, the lack of bright, polarized calibrator sources observable during the daytime, and even at night, the number of well-characterized polarized sources at low radio frequencies remains small. Additionally, ionospheric Faraday rotation can introduce large and rapidly varying polarization angle shifts, further complicating calibration.

A promising approach to address this has recently been proposed by \cite{kansabanik2025ApJS}, where the authors have developed a formalism leveraging the fact that an intrinsically unpolarized sky appears polarized due to instrumental beam polarization. This induced polarization can be modelled and used to derive accurate polarization calibration solutions. We plan to integrate this method into SIMPL, along with solar-specific polarization calibration strategies similar to \citet{kansabanik2022paircarsalgorithm}, to enable spectro-polarimetric snapshot imaging of the Sun with LOFAR.

\subsection{Incorporation of remote baselines}
High-angular-resolution imaging of solar radio emissions is beginning to unveil the fine-scale structure of active regions and offers new avenues to probe coronal scattering and plasma processes \citep[e.g.,][]{mercier2015smallscalestructures,mondal2024ApJ, mondal2025smallspatialstructures, morosan2025resolvingtypeII}. LOFAR, with its remote baselines extending up to $\sim$100 km, holds immense potential for this science. 

The foremost challenge, however, is that the correlated solar flux density drops substantially even before $1000 ~\lambda$, becoming extremely low on longer baselines.
This limitation is compounded by the higher system temperature during solar observations, which further reduces the signal to noise ratio.
Another difficulty arises from ionospheric phase variations, which differ significantly across long baselines, as antennas at larger separations sample different ionospheric patches \citep{lonsdale2005configuration}. 
The self-calibration strategy implemented in SIMPL—based on iterative inclusion of progressively longer baselines—has proven effective for arrays with dense uv-coverage \citep{kansabanik2022workingprinciple}.
However, applying this approach to LOFAR’s remote baselines is ineffective due to their sparse uv-sampling and the associated spatially variable ionospheric distortions. To address this, we are exploring methods to extend \textsc{SIMPL} beyond the core antennas by constructing reliable initial model of the Sun.
One approach involves transferring calibration solutions from night-time calibrator observations, when ionospheric conditions are relatively more stable. 
Under favourable circumstances, these transferred solutions can potentially serve as a starting point for subsequent self-calibration. 
The development and validation of this methodology is currently underway.


\section{Conclusion}\label{sec:conclusion}

We have presented SIMPL, a fully automated, self-calibration-based imaging pipeline developed for LOFAR solar observations. 
This was motivated by the need to address the inadequacies of the standard calibrator-based approaches, arising from the solar contamination of simultaneously observed calibrator sources and the direction-dependent nature of ionospheric phases.  
SIMPL addresses these issues by combining tailored strategies such as optimal calibrator time identification, \textit{uv}-based flagging of solar data, and iterative self-calibration while allowing the model complexity to build up slowly.
This has resulted in an order of magnitude better dynamic range and corresponding improvement in imaging fidelity.
While the current version of SIMPL focuses on Stokes I imaging and excludes remote baselines, future developments aim to extend its capabilities to full-polarization imaging and high-resolution mapping using LOFAR's remote stations. These additions will substantially improve our ability to probe solar radio phenomena across a broader span of spatial and polarization phase space.

Despite their well-recognized science potential \citep[e.g.,][]{cairns2004P&SS, oberoi2004P&SS}, solar radio images remain underutilized by the broader solar physics community. 
This is largely due to the absence of robust, feature-rich analysis toolkits and the steep learning curve associated with adapting generic radio-interferometric tools for solar applications.
SIMPL directly addresses these barriers by delivering reliable high time and frequency resolution spectroscopic snapshot images of the Sun, ready for scientific analysis without requiring specialized expertise in radio interferometry.
The pipeline is now being employed to process over a decade of LOFAR solar interferometric observations, producing science-ready FITS images for open use by the community.
By lowering the technical barriers, SIMPL invites wider participation from the larger solar community and hopes to enable greater integration of solar radio imaging in multi-wavelength studies.

With LOFAR 2.0 on the horizon and the Square Kilometre Array Observatory soon to follow, providing the community access to high-quality science-ready solar radio imaging is more important than ever.
Maximizing the scientific return from these upcoming facilities will require active participation from a diverse, global community transcending traditional discipilnary boundaries, and tools like SIMPL will be key to enable this.

\begin{acknowledgements}
S.D., D.O. and D.P. acknowledge support from the Department of Atomic Energy, under project 12-R\&D-TFR-5.02-0700.
This paper is based on data obtained with the International LOFAR Telescope (ILT). LOFAR \citep{vanhaarlem2013} is the Low Frequency Array designed and constructed by ASTRON. It has observing, data processing, and data storage facilities in several countries, that are owned by various parties (each with their own funding sources), and that are collectively operated by the ILT foundation under a joint scientific policy. The ILT resources have benefitted from the following recent major funding sources: CNRS-INSU, Observatoire de Paris and Université d'Orléans, France; BMBF, MIWF-NRW, MPG, Germany; Science Foundation Ireland (SFI), Department of Business, Enterprise and Innovation (DBEI), Ireland; NWO, The Netherlands; The Science and Technology Facilities Council, UK; Ministry of Science and Higher Education, Poland.
We thank the project LOFAR Data Valorization (LDV) [project numbers 2020.031, 2022.033, and 2024.047] of the research programme Computing Time on National Computer Facilities using SPIDER that is (co-)funded by the Dutch Research Council (NWO), hosted by SURF through the call for proposals of Computing Time on National Computer Facilities. We also thank SURF SARA with the project EINF-13633, Science ready products for LOFAR Solar, Heliospheric and ionospheric datasets.
\end{acknowledgements}

\bibliographystyle{aa}
\bibliography{ref}
\end{document}